\documentclass[aps,prb,reprint,superscriptaddress,
amsmath,citeautoscript,flushbottom,floatfix]{revtex4-1}

\usepackage{graphicx}
\usepackage{color}

\newcommand{\vc}[1]{\boldsymbol{#1}}

\setcitestyle{numbers,square}

\allowdisplaybreaks

\begin{document}
\title{Exchange interactions, Jahn-Teller coupling, and multipole orders \\
in pseudospin one-half $5\boldsymbol{d}^\mathbf{2}$ Mott insulators}

\author{Giniyat Khaliullin}
\email{G.Khaliullin@fkf.mpg.de}
\affiliation{Max Planck Institute for Solid State Research,
Heisenbergstrasse 1, D-70569 Stuttgart, Germany}

\author{Derek Churchill}
\affiliation{Department of Physics, University of Toronto, Ontario, Canada M5S 1A7}

\author{P. Peter Stavropoulos}
\affiliation{Department of Physics, University of Toronto, Ontario, Canada M5S 1A7}

\author{Hae-Young Kee}
\email{hykee@physics.utoronto.ca}
\affiliation{Department of Physics, University of Toronto, Ontario, Canada M5S 1A7}
\affiliation{Canadian Institute for Advanced Research, CIFAR Program in Quantum Materials, Toronto, ON, Canada M5G 1M1}

\begin{abstract}
  We develop a microscopic theory of multipole interactions and orderings in 5$d^2$ transition metal ion compounds. In a cubic environment, the ground state of 5$d^2$ ions is a non-Kramers $E_g$ doublet, which is nonmagnetic but hosts quadrupole and octupole moments. We derive low-energy pseudospin one-half Hamiltonians describing various spin-orbital exchange processes between these ions. Direct overlap of the $t_{2g}$ orbitals results in bond-dependent pseudospin interactions similar to those for $e_g$ orbitals in manganites, except for different orientations of the pseudospin easy axes. On the other hand, the superexchange process, where two different $t_{2g}$ orbitals communicate via oxygen ions, generates new types of pairwise interactions. In perovskites with 180$^\circ$ bonding, we find nearly equal mixture of Heisenberg and $e_g$ orbital compass-type couplings. The 90$^\circ$ superexchange in compounds with edge-shared octahedra is most unusual: despite highly anisotropic shapes of the $E_g$ wavefunctions, the pseudospin interactions have no bond dependence and show instead a hidden SU(2) symmetry, which equally supports quadrupole and octupole orders. We consider the $E_g$ pseudospin models on various lattices and obtain their ground state properties using analytical, classical Monte Carlo, and exact diagonalization methods. On the honeycomb lattice, we observe a duality with the extended Kitaev model, and use it to uncover a critical point where the quadrupole and octupole states are exactly degenerate. On the triangular lattice, an exotic pseudospin state, corresponding to the coherent superposition of vortex-type quadrupole and ferri-type octupole orders, is realized due to geometrical frustration. This state breaks both spatial and time-reversal symmetries, but possesses no dipolar magnetism. We also consider Jahn-Teller coupling effects and lattice mediated interactions between $E_g$ pseudospins, and find that they support quadrupole order. Possible implications of the results for recent experiments on double perovskite osmates are discussed, including effects of local distortions on the pseudospin wavefunctions and interactions.
  
\end{abstract}

\date{\today}

\maketitle


\section{Introduction}

As a hallmark of strong correlations, the spin-orbital multiplet structure of ions is largely preserved in transition metal (TM) compounds. At low temperatures, the spin and orbital degeneracy of these multiplet levels has to be lifted one way or another. Apart from exotic means of the entropy quenching such as formation of quantum spin and orbital liquids, this is typically done by long-range ordering of spins and orbitals, or their composites, through symmetry breaking phase transitions. 

Broadly speaking, the interactions driving these phase transitions have three different microscopic origins: (a) Jahn-Teller orbital-lattice coupling, (b) Kugel-Khomskii type spin-orbital exchange, and (c) relativistic spin-orbit coupling (SOC). Depending on the multiplet structure of constituent ions and the nature of chemical bonds in a crystal, the interplay between these couplings may take various forms, resulting in rich spin-orbital physics in TM compounds. 

Lifting the orbital degeneracy via a cooperative Jahn-Teller (JT) structural transition is most common in $e_g$ orbital systems like manganites. At this transition, the orbitals are (self)trapped by static lattice distortions. The JT driven orbital order is essentially independent of spins and happens well before magnetic ordering. In this picture, the low energy physics is given by ``spin-only'' Hamiltonians, with the exchange parameters dictated by the Goodenough-Kanamori rules~\cite{Goo63,Kug82}.

In $t_{2g}$ orbital systems with relatively weak JT coupling, the spin $\vc S$ and orbital $\vc L$ degrees of freedom are no longer separated~\cite{Kha05}. They may instead develop joint dynamics driven by the spin-orbital exchange interactions, as well as by intraionic SOC which unifies the two sectors by forming total angular momentum $\vc J=\vc S+\vc L$. The latter root to the ``spin-orbital-entangled'' physics is especially relevant to 4$d$ and 5$d$ electron compounds. 

In the strong SOC limit, the JT orbital-lattice coupling and Kugel-Khomskii exchange interactions have to be reformulated in terms of total angular momentum $\vc J$ of the lowest multiplet level, as is usually done in 4$f$ electron systems. This leads to a number of important consequences. First, the JT orbital order is ``converted'' into quadrupole order of $\vc J$ moments, involving also the spin sector which was initially ``blind'' to JT physics. Effective JT coupling is typically reduced, due to a partial suppression of the initial orbital degeneracy. Second, exchange interactions between effective $\vc J$ moments (``pseudospins'') may become highly anisotropic and bond-directional; this is due to the non-spherical shape of the spin-orbit entangled wavefunctions. Third, pseudospin states may carry not only dipole or quadrupole moments, but also higher-rank multipoles such as a magnetic octupole. 

The physical content of pseudospin wavefunctions is decided by a filling factor $n$ of $d$-orbital levels. In combination with the lattice and chemical bonding geometry in a given material, this leads to a variety of non-trivial interactions and ground states among different $d^n$ compounds. This includes a possible realization of Kitaev spin-liquids, excitonic magnetism, and multipole orders (for a recent review, see Ref.~\cite{Tak21}). 

In this paper, we focus on spin-orbital physics in compounds based on $d^2$ ions. The $d^2$ configuration with two-electron spin $S=1$ and effective orbital moment $L=1$ is special, because its total angular momentum $J=2$ is isomorphic to a single $d$-electron orbital moment, $l=2$. This analogy has interesting implications for the symmetry and physical properties of $d^2$ ions. Namely, in a cubic environment, a $J=2$ level has to split into $E_g$ doublet and $T_{2g}$ triplet levels [see Fig.~1(a)], just like the $d$-electron $l=2$ level splits into $e_g$ and $t_{2g}$ orbital levels~\cite{note_G3}. While the $T_{2g}$ triplet hosts an effective angular momentum $\widetilde{J} =1$ (with a familiar relation $\widetilde{\vc J}=-\vc J$)~\cite{Abr70}, the non-Kramers $E_g$ doublet is similar to an $e_g$ doublet and carriers no dipole moment. This implies that $d^2$ ions with non-Kramers $E_g$ ground states may show high-rank multipole orders similar to rare-earth $f^2$ non-Kramers $\Gamma_3$ ions~\cite{Kur09}. 

Experimentally, a single phase transition around 30-50~K is observed in 5$d^2$ double perovskite (DP) compounds~\cite{Mah20,Mar16,Tho14,Aha10}. This is very different from 5$d^1$ Kramers ion DPs which show two separate transitions~\cite{Lu17,Hir19,Hir20,Hir21}, corresponding to quadrupole (structural) and dipole orders of $J=3/2$ states~\cite{Che10,Wit14}. Having a single transition is natural for pseudospin-1/2 doublet systems, and this clearly points to the $E_g$ doublet physics in 5$d^2$ DPs. However, the precise nature of this transition is not yet fully established. The structural changes at this transition, if any, are found to be below 0.1\%~\cite{Mah20}. While no magnetic Bragg peaks were seen in neutron diffraction data, time-reversal (TR) symmetry breaking is detected by muon spin relaxation. To reconcile these observations, a ferro-type octupolar order of the $E_g$ doublets has been proposed~\cite{Mah20,Par20,Vol20}. 

The octupole is a third-rank magnetic multipole which carries no dipole moment, and its long-range order is observed in rare-earth compounds (see Ref.~\cite{Kur09} for a review of multipole orders). The possibility of octupolar order in $d$ electron systems is intriguing. It is actually quite unexpected because an $E_g$ doublet is subject to JT physics: its partners have different charge density shapes (planar and elongated), see Fig.~1(a). Therefore, a conventional quadrupole order like in $e_g$ orbital systems~\cite{Kug82} is the most natural instability to expect in the first place. To realize the octupolar order, exchange interactions between the octupole moments must be strong enough to overcome the quadrupolar interactions contributed by the Kugel-Khomskii exchange and orbital-lattice JT couplings.

Early theoretical work~\cite{Che11} on $d^2$ DP systems with strong SOC assumed that cubic splitting of the $J=2$ level $\Delta_c$ is smaller than the exchange couplings and therefore neglected it. The obtained phase diagram contains dipolar and quadrupolar ordered states. Here we develop a theory of $d^2$ electron systems starting from the opposite limit, i.e. when cubic splitting $\Delta_c$ is large and the $E_g$ doublet is well separated from the virtual $T_{2g}$ states, as actually seen in experiment~\cite{Mah20}. Having in mind 5$d^2$ materials other than DP compounds, we keep the discussion as general as possible, considering various spin-orbital exchange processes typical in TM oxides. The resulting $E_g$ doublet interactions are represented in terms of pseudospin one-half Hamiltonians. In most cases, the interactions are dominated by quadrupolar couplings. In a 90$^\circ$ exchange geometry however, the quadrupole and octupole channels are equally present, and effective interactions on a single bond can be written in a Heisenberg form with no preference for either of these two channels. The resulting multipole orders of $E_g$ doublets in different lattices are considered. On a honeycomb lattice, we show that the $E_g$ pseudospin model can be mapped to the extended Kitaev model, thereby uncovering a hidden SU(2) symmetry point that separates quadrupole and octupole orders. The pseudospins on a geometrically frustrated triangular lattice show more complex phase behavior, including a coherent mixture of different rank (quadrupole and octupole) orders in the ground state. The order parameters are reduced by quantum fluctuations. In DP lattices, we find that the exchange interactions favor a quadrupole order.

We further discuss orbital-lattice coupling effects, and show that JT phonon mediated interactions cooperate with exchange interactions to support quadrupolar order. This is similar to conventional $e_g$ orbital systems. We suggest that in DP lattices, where the magnetic ions are widely separated and have no common oxygen, a dynamical Jahn-Teller effect may develop to reduce the structural distortions induced by quadrupolar order. We also consider modifications of the pseudospin wavefunctions by symmetry lowering distortions (caused by site disorder or other defects), and find that they induce a magnetic dipole moment on the $E_g$ doublet. In general, $d^2$ compounds represent an interesting class of materials where all three main actors - the electron exchange, orbital-lattice interaction, and relativistic SOC - play an essential role in determining the ground states and low-energy excitations.   

The paper is organized as follows: Sec.~\ref{sec:Eg} introduces the $E_g$ doublet states and their pseudospin-1/2 description. In Section~\ref{sec:Ham}, we derive pseudospin Hamiltonians considering different orbital exchange geometries which are typical in TM compounds. Sec.~\ref{sec:Ord} studies pseudospin orderings and excitations on various lattice structures. Sec.~\ref{sec:JT} discusses Jahn-Teller coupling and disorder effects in the context of experiments in DP compounds. Sec.~\ref{sec:Con} summarizes the main results.


\section{Non-Kramers $E_g$ doublet and pseudospins}
\label{sec:Eg}

The $E_g$ doublet wavefunctions written in the $J_z$ basis are \cite{Abr70}:
$\tfrac{1}{\sqrt 2}(|2\rangle+|\!-2\rangle)$ and $|0\rangle$. We regard them as pseudospin $s=1/2$ states $|\!\uparrow\rangle$ and $|\!\downarrow\rangle$, correspondingly. To get an idea about the orbital shapes, one can represent these functions in terms of two-electron spin and orbital $|S_z, L_z\rangle$ states: 
\begin{align}
|\!\uparrow\rangle &=\tfrac{1}{\sqrt 2}(|1,1\rangle+|\!-1,-1\rangle), \\
|\!\downarrow\rangle &=\tfrac{1}{\sqrt 6}(|1,-1\rangle+2|0,0\rangle+|\!-1,1\rangle).
\label{eq:spin}
\end{align}
In the pseudospin-up state with $L_z=\pm 1$, one of the electrons must occupy $l_z=0$ planar orbital $d_{xy}$, flattening the overall charge density as shown in Fig.~1(a). However, the pseudospin-down state is dominated by an $L_z=0$ component, where the electrons occupy $l_z=1$ and $l_z=-1$ complex orbitals $\mp(d_{yz}\pm id_{zx})/\sqrt{2}$; thus, its charge density is elongated towards apical oxygen $O_z$. Under cubic rotations, these wavefunctions transform in a standard way, similar to $e_g$ orbital pair, $x^2-y^2$ and $3z^2-r^2$.

Within the $E_g$ doublet, the $J=2$ quadrupole operators 
\begin{align}
\label{eq:O3}
O_3 &= \frac{1}{6}(2J_z^2-J_x^2-J_y^2), \\
O_2 &= \frac{1}{2\sqrt 3}(J_x^2-J_y^2)  
\label{eq:O2}
\end{align}
have matrix elements $\langle \pm\frac12|O_3|\pm\frac12\rangle=\pm 1$ and 
$\langle \pm\frac12|O_2|\mp\frac12\rangle=1$. Thus, the following correspondence between the pseudospin $s^z$ and $s^x$ components, and $E_g$ quadrupoles follows: $s^z=\frac12 O_3$ and $s^x=\frac12 O_2$. The third component $s^y=\frac12 T_{xyz}$ describes the octupolar moment $T_{xyz}=\frac{1}{\sqrt 3} \overline{J_xJ_yJ_z}$ with threefold symmetry axis [111]. The projections of the octahedral $x,y,z$ axes onto the two-dimensional pseudospin $(s^z,s^x)$ plane [111] make 120$^\circ$ angles between them, and the pseudospin $s^z$ axis is parallel to the octahedral $z$ axis projection, see Fig.~1(a). This is the most natural choice, because $s^z$ is related to the $O_3$ quadrupole moment (\ref{eq:O3}) directed along $z$ axis. As we will see below, this also results in one-to-one correspondence between the exchange bond labels $\gamma \in\{x,y,z\}$ and octahedral $(x,y,z)$ axes. The basis rotations within the $(s^z,s^x)$ plane by $\phi=2\pi/3$ correspond to the cyclic permutations among $J_x,J_y,J_z$. Finally, we note that the $s^z$ and $s^x$ operators are TR-even, while the $s^y$ octupole is TR-odd; this implies that the pairwise interactions of the type $s^z_is^y_j$ and $s^x_is^y_j$ are not allowed, unless TR symmetry is broken. 

Following $e_g$ orbital pseudospin formalism \cite{Kan60,Kug82}, we introduce the following pseudospin combinations:
\begin{align}
\label{eq:tauz} 
\tau_\gamma &=\;\;\;\cos\phi_\gamma \;s^z+\sin\phi_\gamma \;s^x, \\ 
\bar\tau_\gamma &=-\sin\phi_\gamma \;s^z+\cos\phi_\gamma \;s^x. 
\label{eq:taux}
\end{align}
Here, the pseudospin index $\gamma=(z,x,y)$ also specifies the corresponding angles $\phi_\gamma=(0, 2\pi/3, 4\pi/3)$. In essence, $(\tau_\gamma,\bar\tau_\gamma)$ play the role of $(s_\gamma^z,s_\gamma^x)$ operators defined in the rotated basis of pseudospin functions:
\begin{align}
  |\!\uparrow\rangle_\gamma &=\;\;\;\cos(\phi_\gamma/2)|\!\uparrow\rangle +
  \sin(\phi_\gamma/2)|\!\downarrow\rangle, \\ 
|\!\downarrow\rangle_\gamma &=-\sin(\phi_\gamma/2)|\!\uparrow\rangle +\cos(\phi_\gamma/2)|\!\downarrow\rangle. 
\label{eq:f_phi}
\end{align}
Physically, $\tau_x$ ($\tau_y$) and $\bar\tau_x$ ($\bar\tau_y$) correspond to the quadrupolar operators of $3x^2-r^2$ ($3y^2-r^2$) and $y^2-z^2$ ($z^2-x^2$) symmetries, respectively. The notations $\tau_\gamma$ and $\bar\tau_\gamma$ are useful since one may derive the exchange Hamiltonian $\mathcal{H}^{(\gamma)}$ for $\gamma=z$ type bonds in terms of $(s^z, s^x)$ pair, and then restore $\mathcal{H}^{(\gamma)}$ for all $\gamma$ by simply replacing $s^z \rightarrow \tau_\gamma$ and $s^x \rightarrow \bar\tau_\gamma$. In perovskites with 180$^\circ$ bonding, the $z$-type bond is parallel to the octahedral $z$ axis; while in other cases, e.g. in a honeycomb lattice, the $z$-type bond is orthogonal to the octahedral $z$ axis (a convention also used in the Kitaev model literature). 

To derive pseudospin exchange interactions, one has to project Kugel-Khomskii type spin-orbital Hamiltonians -- which are already known from previous works -- onto the low-energy $E_g$ doublet subspace. We note that conventional $e_g$ orbital exchange interactions operate in the quadrupolar sector $(s^z,s^x)$ exclusively \cite{Kug82}. In contrast, we will see below that the $E_g$ ``orbital'' exchange may involve interactions between the octupole moments $s^y$, as well; this is because the $E_g$ pseudospin states are spin-orbit entangled objects. Combined with the specific hopping geometry of $t_{2g}$ orbitals, this results in a nontrivial structure of the $E_g$ interactions. We consider below some basic exchange processes which commonly appear in transition metal compounds.


\begin{figure}[tb]
\begin{center}
\includegraphics[scale=0.4]{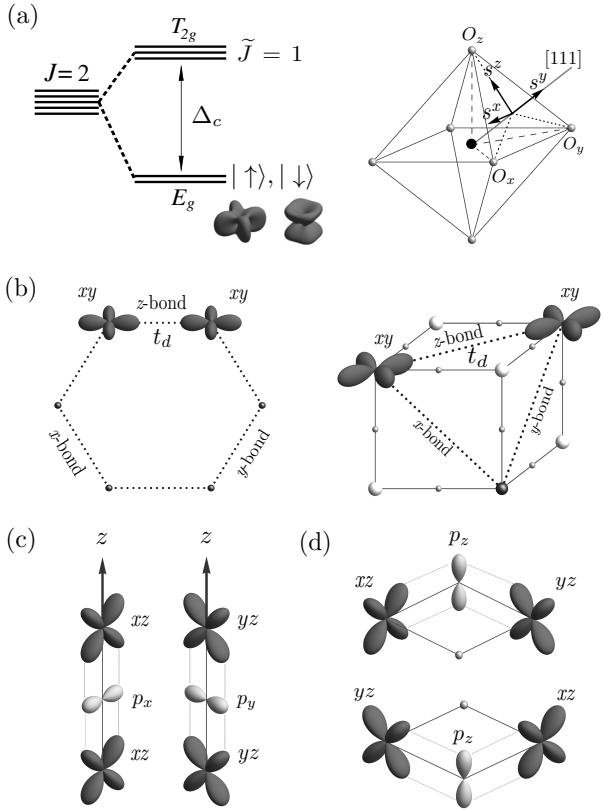}
\caption{
(a)~Cubic splitting $\Delta_c$ of the $J=2$ level, and the spatial shapes of $E_g$ doublet wavefunctions. Right panel shows the pseudospin ($s^x,s^y,s^z$) coordinate axes with respect to oxygen octahedra.    
(b)~Direct hopping between $xy$ orbitals in honeycomb (left) and DP (right) lattices, resulting in bond-dependent pseudospin $\tau$-interactions Eq. (\ref{eq:tau}) between spin-orbit entangled $E_g$ states. 
(c)~Two-orbital superexchange via 180$^\circ$ Me-O-Me bonding geometry. Hopping is orbital conserving: $xz \leftrightarrow xz$ and $yz \leftrightarrow yz$. This process results in pseudospin interactions Eq. (\ref{eq:180}), comprising isotropic Heisenberg and bond-dependent compass-type couplings. 
(d)~Two-orbital superexchange via 90$^\circ$ bonding geometry. Hopping interchanges the orbital labels: $xz \leftrightarrow yz$. This process leads to the interactions Eq. (\ref{eq:90}), which are anisotropic in pseudospin space but have no bond dependence.
}\label{fig:shapes}
\end{center}
\end{figure}

\section{Pseudospin exchange Hamiltonians}
\label{sec:Ham}

\subsection{Single-orbital exchange: direct $t_{2g}$ orbital overlap}
\label{sec:dd}

We start with the simple case where one specific orbital is active on a given nearest-neighbor (NN) exchange bond. Two examples of single-orbital coupling are shown in Fig.~1(b): direct $d_{xy}$ orbital hopping on $z$-type bonds in the honeycomb lattice, and $d_{xy}$ orbital hopping in the $ab$ plane of the DP lattice \cite{Che11}. In this case, we expect that the $E_g$ exchange Hamiltonian is similar to that for $e_g$ orbitals in ferromagnetic manganites. Indeed, spin-orbit $E_g$ and pure orbital $e_g$ states have the same ($\Gamma_3$) symmetry properties. Moreover, the Kugel-Khomskii $e_g$ exchange process also involves a single-orbital, specific to a given bond \cite{Kug82}. 

Neglecting Hund's coupling effects in the intermediate states, direct hopping $-t_d(d_{xy,i}^\dagger d_{xy,j}^{\phantom{\dagger}}+\mathrm{H.c.})$ gives the following exchange Hamiltonian, written in terms of spin $S=1$ and orbital $L=1$ moments of a $d^2$ ion \cite{Cha19}:
\begin{equation}
\mathcal{H}_{ij}^{(c)} = \frac{t_d^2}{U}\,\left[(\vc S_i\cdot\vc S_j+1)
  L_{zi}^2L_{zj}^2\!-\!L_{zi}^2\!-\!L_{zj}^2\right].
\label{eq:Hdd}
\end{equation}
For the $x$ ($y$) bonds where the $d_{yz}$ ($d_{zx}$) orbital exchange is active, $L_z$ is replaced by $L_x$ ($L_y$). Projection of this Hamiltonian onto the $E_g$ subspace results in: 
\begin{equation}
\mathcal{H}^{(\gamma)}_{ij}(d) =J_\tau \; \tau_{i\gamma}\tau_{j\gamma} \;,
\label{eq:tau}
\end{equation}
with $J_\tau=\frac{4}{9}\frac{t_d^2}{U}$, and $\tau_\gamma$ given by Eq. (\ref{eq:tauz}). This interaction has the same structure as the Kugel-Khomskii $e_g$ orbital Hamiltonian, but with the reduced exchange constant due to a small fraction of the active orbital (e.g. $d_{xy}$ for the $z$ bond) in the two-electron $E_g$ wavefunction. Representative values of $t_d\sim 0.1-0.2$~eV and $U\sim 2$ eV would give an energy scale of $J_\tau\sim 2-9$~meV (the lower end is appropriate for DP lattice where $d$-ions are well separated and thus hopping $t$ is small).

We note that there is a subtle difference between Eq.~(\ref{eq:tau}) and the Kugel-Khomskii $e_g$ exchange in perovskites~\cite{Kug82}. In the latter, the active $e_g$-orbital (say $3z^2-r^2$ on $z$ bond) is quasi-one dimensional and bond-oriented, thus enforcing pseudospins $\tau_\gamma$ to be aligned along the interacting $\gamma$-bond directions (hence the name ``pseudodipolar'' or ''compass'' model~\cite{Kug82}). In contrast, the $\tau_\gamma$ quadrupoles in Eq.~(\ref{eq:tau}) try to avoid the bond directions; e.g. for $z$-type bond we have $s_i^zs_j^z$ coupling but with the Ising $s^z$ axis being perpendicular to the $z$ bond direction. Physically, the pseudospin orientation specifies the shape of the quadrupolar charge distribution and can be probed in the experiment. Formally however, the two models can be converted into each other by a 90$^\circ$ rotation within the ($s^x,s^z$) quadrupolar plane [i.e. replacing $\tau$ in Eq.~(\ref{eq:tau}) by $\bar\tau$]. This point has to be kept in mind while comparing the present $\tau$-model results with those in canonical compass model studies~\cite{Nas08,Zha08,Wu08,Che16,Nus15}. 

The bond-dependent nature of the interactions in Eq.~(\ref{eq:tau}) brings about frustration effects intrinsic to Kugel-Khomskii type spin-orbital models~\cite{Kug82} and their descendants~\cite{Kha05,Nus15}. Typically, this frustration is resolved by order-from-disorder mechanism, see, e.g. Refs.~\cite{Kha01,Mos02,Kub02}. 


\subsection{Two-orbital superexchange: 180$^\circ$ bonding geometry}
\label{sec:180}

This case is typical for a metal-oxygen-metal (Me-O-Me) superexchange process in perovskites, see Fig. 1(c). On the $z$-type bond, two orbitals $a=d_{yz}$ and $b=d_{zx}$ equally contribute, and hopping is orbital-conserving: $-t(a_{i\sigma}^\dagger a_{j\sigma}^{\phantom{\dagger}} + b_{i\sigma}^\dagger b_{j\sigma}^{\phantom{\dagger}})$. The spin-orbital Hamiltonian (at $J_H=0$) reads as \cite{Kha13}:
\begin{equation}
\mathcal{H}^{(\gamma)}_{ij}=\frac{t^2}{U}[(\vc{S_i}\cdot\vc{S_j}+1)\mathcal{O}_{ij}^{(\gamma)}+
  (L_i^\gamma)^2+(L_j^\gamma)^2],
\label{eq:HSL}
\end{equation}
where orbital operator for $z$ type bond reads as 
\begin{equation}
\mathcal{O}_{ij}^{(z)}=(L_i^xL_j^x)^2+(L_i^yL_j^y)^2+L_i^xL_i^yL_j^yL_j^x+L_i^yL_i^xL_j^xL_j^y.
\label{eq:H180}
\end{equation}
Operators $\mathcal{O}^{(x)}$ and $\mathcal{O}^{(y)}$ for $x$ and $y$ bonds follow from cubic permutations among $L_x,L_y,L_z$.

A projection of the above Hamiltonian onto the $E_g$ subspace gives
\begin{equation}
  \mathcal{H}^{(\gamma)}_{ij}(180^\circ) =J  \left[\vc{s_i}\cdot\vc{s_j} +
  \tfrac{2}{3}\tau_{i\gamma}\tau_{j\gamma}\right],
\label{eq:180}
\end{equation}
with the exchange constant $J=\frac{2}{3}\frac{t^2}{U}$. In this equation, the Heisenberg term gives an equal coupling in quadrupolar $(s^x, s^z)$ and octupolar $s^y$ sectors. This term is not present in the Kugel-Khomskii $e_g$ orbital Hamiltonian, but is realized here because the $E_g$ orbitals have a complex internal structure, and they are made of $t_{2g}$ orbitals with hopping rules different from those for $e_g$ orbitals.

The second bond-dependent $\tau_\gamma$ term in Eq. (\ref{eq:180}) is a direct analogue of the Kugel-Khomskii $e_g$ orbital exchange. It operates only in the quadrupolar channel, thus disfavoring octupolar correlations. It is important to note that the easy axis orientations in this term exactly coincide with the bond directions, as dictated by the shapes of the active complex orbitals, e.g. $\mp(d_{yz}\pm id_{zx})/\sqrt{2}$ orbitals having rotational symmetry around $z$ bond (like $3z^2-r^2$ axial symmetry in $e_g$ models). Thus the $E_g$ pseudospins in the 180$^\circ$ bonding geometry behave exactly as the $e_g$ orbital compasses do in cubic lattices \cite{Kug82}, orienting themselves along the bond directions. This follows from a general observation that in case of axial symmetry, the spin-1/2 anisotropy term should have a dipole-dipole interaction form \cite{Van37}. For the same reason, the compass-like $\tau$-interaction also appears for non-Kramers doublets in $f^2$ electron system \cite{Kub17}. Due to differences between $d$ and $f$ orbital hopping geometries however, the isotropic term in Eq.~(\ref{eq:180}) is not present in the $f^2$ case. On square or cubic lattices, we expect that the Hamiltonian (\ref{eq:180}) would have two-sublattice quadrupolar order, with alternating planar and elongated $E_g$ states, as selected by the anisotropic $\tau$-term via order-from-disorder mechanism. 


\subsection{Two-orbital superexchange: 90$^\circ$ bonding geometry}
\label{sec:90}

This process is typical for nearest-neighbor Me-O$_2$-Me superexchange in delafossite derived structures with edge shared octahedra, see Fig.~1(d). On the $z$ type bond, two orbitals $a=d_{yz}$ and $b=d_{zx}$ equally contribute again, but hopping is orbital non-conserving: $-t(a_{i\sigma}^\dagger b_{j\sigma}^{\phantom{\dagger}} + b_{i\sigma}^\dagger a_{j\sigma}^{\phantom{\dagger}})$. The resulting spin-orbital Hamiltonian reads as in the 180$^\circ$ case, see Eq.~(\ref{eq:HSL}), but now with the modified orbital part [i.e. interchanging $L_j^x \leftrightarrow L_j^y$ in Eq.~(\ref{eq:H180})] \cite{Kha13}:
\begin{equation}
\mathcal{O}_{ij}^{(z)}=(L_i^xL_j^y)^2+(L_i^yL_j^x)^2+L_i^xL_i^yL_j^xL_j^y+L_i^yL_i^xL_j^yL_j^x.
\label{eq:H90}
\end{equation}
The orbital non-conservation during the hoppings has dramatic consequences for the exchange symmetry, as observed previously in spin-orbit $J=1/2$ \cite{Kha05,Jac09} and $J=0$ \cite{Kha13} systems. In the present non-Kramers $E_g$ doublet case, this results in the pseudospin Hamiltonian:
\begin{equation}
\mathcal{H}_{ij}(90^\circ) =
J \; (s_i^ys_j^y - s_i^xs_j^x - s_i^zs_j^z),
\label{eq:90}
\end{equation}
which is completely different from $\mathcal{H}_{ij}(180^\circ)$ in Eq. (\ref{eq:180}), but the coupling constant remains the same: $J=\frac{2}{3}\frac{t^2}{U}$. This result is remarkable in several aspects. It has no $\gamma$-bond dependence, since the quadrupolar ($s^x, s^z$) part is isotropic and does not change under the rotations (\ref{eq:tauz}) and (\ref{eq:taux}), and the octupolar $s^y$ moment is not affected by $C_3$ rotations around [111] axis and is thus independent of $\gamma$ as well. This is unlike the $d^5$ Kramers doublet case, where the cubic rotations affect all three components of the $J=1/2$ vector, via cyclic permutations of its $x, y, z$ components (see, e.g. Eq. (5.8) in Ref.~\cite{Kha05}). Nevertheless, SOC results in strong exchange anisotropy: quadrupoles are ferro-correlated, while the octupolar components $s^y$ are coupled in an antiferro-fashion. 

In bipartite (e.g. honeycomb) lattices, this anisotropic Hamiltonian can conveniently be converted into an AF Heisenberg form $J \vc{s_i}\cdot\vc{s_j}$, by changing the sign of the $s^x$ and $s^z$ components on one of the sublattices; such hidden symmetries are common to spin-orbit pseudospin-1/2 Hamiltonians \cite{Kha05,Cha10,Cha15}. After this transformation, one observes an exact degeneracy between quadrupole and octupolar orderings, with the out-of-plane Goldstone mode representing a smooth rotation from one type order to the other one at no energy cost. Such exact degeneracy and coherent mixture of different (even/odd) rank order parameters and related gapless modes is rather unusual, but have previously been discussed in the context of $t_{2g}$ orbital Hamiltonians, see Refs.~\cite{Kha03,Kha05} for details.


\section{Pseudospin order: Quadrupolar versus octupolar states} 
\label{sec:Ord}

In this section, we discuss the phase behavior and excitations of the $E_g$ pseudospin models on different lattices.

\subsection{Simple cubic lattice}
\label{sec:cub}

The two-orbital 180$^\circ$-exchange Hamiltonian, Eq.~(\ref{eq:180}), is applicable to perovskite lattices. As we already mentioned in that section, we expect a two-sublattice quadrupolar order in this case. This is conceptually similar to $e_g$ orbital order in 3$d$ systems; the only difference is that the $E_g$ ``orbitals'' are spin-orbit coupled objects. Like in the $e_g$ orbital case, both exchange and JT couplings will contribute to the quadrupolar ordering, and they typically support each other. Formally, the Hamiltonian (\ref{eq:180}), comprising an AF Heisenberg interaction and an anisotropic compass-like terms is very similar to the model studied in Ref.~\cite{Kha01}. Thus, its excitation spectrum should acquire a sizeable gap due to the order-by-disorder mechanism.

\subsection{Honeycomb lattice}
\label{sec:hon}

\begin{figure}[tb]
\begin{center}
\includegraphics[width=0.49\textwidth]{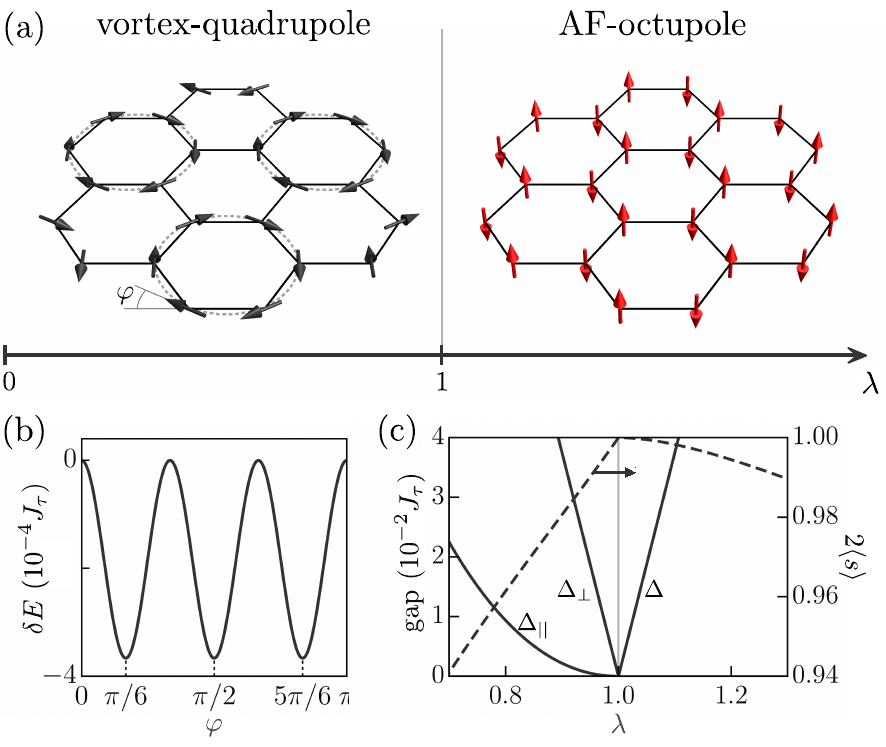}
\caption{
(a)~Phase diagram of the Hamiltonian (\ref{eq:KH}) on honeycomb lattice as a function of $\lambda=2J/J_\tau$. There are two states, separated by a first-order spin-flop transition at the hidden SU(2) symmetric point $\lambda =1$. Left and right insets show the ordered patterns of the vortex-type quadrupole and AF-octupole phases, respectively. 
(b)~Quantum zero-point energy $\delta E$ as a function of ordered moment orientation $\varphi$ within the honeycomb plane, calculated at $\lambda = 0.5$. A ground state pattern with $\varphi =\pi/6$ is shown in panel (a) left. 
(c)~The in-plane ($\Delta_{\parallel}$) and out-of-plane ($\Delta_{\perp}$) magnon gaps in the vortex-quadrupole phase, and the magnon gap ($\Delta$) in the uniaxial AF-octupole ordered state. Dashed line shows the pseudospin order parameter length 2$\langle s \rangle$ near the transition point, where the model is dual to a fluctuation free Heisenberg FM.  
}\label{fig:honey}
\end{center}
\end{figure}

A honeycomb lattice is derived from the delafossite structure with edge-shared octahedra. In general, two different channels are operative in this case: direct hopping $t_d$ considered in Sec.~\ref{sec:dd}, and indirect $t$ superexchange via 90$^\circ$ bonding considered in Sec.~\ref{sec:90}. It is known that for pseudospin $J=1/2$ exchange in $d^5$ compounds, there is also a combination of these two processes (i.e. $t$ times $t_d$ terms) resulting in the off-diagonal, so-called $\Gamma$ interaction \cite{Rau14}. Interestingly, such a cross-term is absent in the present $E_g$ problem. So, the full Hamiltonian in honeycomb or triangular lattices is comprised of the bond-dependent $\tau$-model (\ref{eq:tau}), and the 90$^\circ$ exchange $J$-interaction (\ref{eq:90}) which is also anisotropic, but bond-independent: 
\begin{equation}
  \mathcal{H}_{ij}^{(\gamma)}=J_\tau \; \tau_{i\gamma}\tau_{j\gamma}
  +\; J \;(s_i^ys_j^y - s_i^xs_j^x - s_i^zs_j^z).
\label{eq:JtauJ}
\end{equation}

Physically, both $J_\tau$ and $J$ are positive, and their ratio can be arbitrary. While the first term operates in the pure quadrupolar $(s^z,s^x)$ sector, $J$ coupling equally supports AF octupolar and FM quadrupolar states. As noticed above, the $J$ interaction is actually dual to the AF Heisenberg model (on bipartite lattices). A finite $\tau$-term breaks this symmetry and selects the ordering type. Since the $(s^z,s^x)$ part of the $J$ term has a negative sign, a small admixture of positive $J_\tau$ reduces the quadrupole interactions. As a result, two-sublattice staggered order of octupole moments $s^y$ is favored at $J \gg J_\tau$. The ground state wavefunction is complex, $\psi_{A/B}=(|\!\!\uparrow\rangle \pm i|\!\!\downarrow\rangle)/\sqrt 2$, and has a cubic shape, see Fig.~2(c) of Ref.~\cite{Tak21}. 


In the opposite limit $J_\tau \gg J$, it is obvious that $s^y$ octupole order has to give way to ordering of the $\tau$-quadrupoles that live in $(s^z,s^x)$ plane. The quadrupole order is TR invariant (i.e. the condensate wavefunction is real) but breaks cubic symmetry. The transition is of a spin-flop type: spins flop from the [111] direction into the honeycomb plane. In terms of the condensate wavefunction, this corresponds to the phase-jump from $\pi/2$ to $0$ in the relative phase factor $e^{i\phi}$ between $|\!\uparrow\rangle$ and $|\!\downarrow\rangle$ states. 

Interestingly, the transition point and quadrupole order pattern that replaces octupole order can be obtained from symmetry considerations alone, by virtue of the duality transformations in pseudospin honeycomb models~\cite{Cha15}. To this end, we use the explicit form of $\tau_\gamma$ given in Eq.~(\ref{eq:tauz}) and re-write Eq.~(\ref{eq:JtauJ}) as follows:
\begin{align} \nonumber
\mathcal{H}_{ij}^{(\gamma)}&=
(1-\lambda)(s_i^zs_j^z+s_i^xs_j^x)\;+\;\lambda s_i^ys_j^y \\
&+ \cos\phi_\gamma (s_i^zs_j^z-s_i^xs_j^x)-\sin\phi_\gamma (s_i^zs_j^x + s_i^xs_j^z). 
\label{eq:KH}
\end{align}
Here, $\lambda=2J/J_\tau$, and the overall energy scale equal to $J_\tau/2$ is not shown. This equation has exactly the same structure as the extended Kitaev model, written in the hexagonal coordinate frame~\cite{Cha15}. Simple re-labeling of the spin axes $(x,y,z) \leftrightarrow (Y,Z,X)$, and a term-by-term comparison of Eq.~(\ref{eq:KH}) with Eq.~(A1) of Ref.~\cite{Cha15} gives the following correspondence: $J_{XY}=1-\lambda$, $J_Z=\lambda$, $A=1$, and $B=0$. (We note that $B$ term of Ref.~\cite{Cha15} couples in-plane and out-of-plane components of spins; for the present $E_g$ problem, finite $B$ would imply linear quadrupole-octupole coupling which is forbidden by TR symmetry). Next, we use the relations (A2-A5) of Ref.~\cite{Cha15} to obtain the parameters $K$, $\Gamma$, $\bar J$, and $\Gamma'$, which define the extended Kitaev model in the octahedral axes frame~\cite{Rau14} (we use $\bar J$ to avoid confusion with $J$ in our models):
\begin{align} 
K &= 1,                            \label{eq:K} \\
\Gamma &= 1- \frac23 (1-\lambda),  \label{eq:G} \\
\bar J &= \frac13 (1-\lambda),     \label{eq:J} \\
\Gamma' &= - \frac23 (1-\lambda).  \label{eq:G'}
\end{align}

So far, we have shown that Eqs.~(\ref{eq:JtauJ}) and (\ref{eq:KH}) correspond to the extended Kitaev model at the specific parameter set. The virtue of this mapping is that at $\lambda=2J/J_\tau=1$, we see that $\bar J = \Gamma'=0$. Thus, at this point, the model is isomorphic to the $K=\Gamma=1$ model, which in turn, is dual to the isotropic Heisenberg model, see Table I of Ref.~\cite{Cha15}. This leads to a remarkable observation that at $J_\tau=2J$, the highly anisotropic Hamiltonian (\ref{eq:JtauJ}) is dual to the effective FM Heisenberg model $\mathcal{\tilde H}_{ij}=-J\vc{\tilde s_i}\cdot\vc{\tilde s_j}$. The duality transformation involves a six-sublattice rotation matrix $\mathcal{T}_6$~\cite{Cha15}, which converts in-plane FM order of effective spins $\vc{\tilde s}$ into a vortex pattern of $s^z$ and $s^x$ moments in our model. This quadrupole order is shown in Fig.~\ref{fig:honey}(a) (cf. Fig.~2(e) of Ref.~\cite{Cha15}). On the other hand, out-of-plane FM order of $\vc{\tilde s}$ corresponds to octupolar AF order of $s^y$ moments already discussed above. Being dual to the eigenstates of a hidden FM Heisenberg model, these vortex and AF states are ``fluctuation free'', and low-energy excitations are magnons with a quadratic dispersion. 

The exact degeneracy of these two states is lifted as soon as $\lambda$ deviates from its critical value 1. From Eq.~(\ref{eq:KH}), we see that the corrections to the SU(2) point Hamiltonian $\mathcal{H}(\lambda\!\!=\!\!1)$ read as 
$(1-\lambda)(s_i^zs_j^z +s_i^xs_j^x-s_i^ys_j^y)$. This term acts as an easy-plane or easy-axis anisotropy, selecting quadrupolar vortex order if $1-\lambda>0$, and $s^y$ octupolar AF order if $1-\lambda<0$. It also opens a finite gap in magnon spectra. 

The symmetry-based considerations above are confirmed by numerical studies. Fig.~\ref{fig:honey}(a) shows the phase diagram of Hamiltonian (\ref{eq:KH}), obtained by classical theory and exact diagonalization (ED) on a C$_3$ symmetric 24-site cluster. The first-order transition between the vortex-quadrupole and AF-octupole phases occurs at $\lambda = 1$. Across the spin-flop transition, the in-plane quadrupole moments (black arrows) flip to the out-of-plane octupole moments (red arrows). We investigated the quantum effects using linear spin wave theory (LSWT). In the quadrupole phase, the zero-point magnon energy $\delta E$ depends on the vortex pattern orientation [specified by the angle $\varphi$ in Fig.~\ref{fig:honey}(a)]. The state with $\varphi = \pi/6$ has the lowest energy, see Fig.~\ref{fig:honey}(b). Note that the pinning potential is extremely weak, so the spins are almost free to rotate (globally) within a quadrupolar plane. This implies the presence of low-energy quadrupole moment fluctuations.   

The calculated magnon gaps are presented in Fig.~\ref{fig:honey}(c) near the spin-flop transition area ($\lambda \sim 1$). In the planar-type quadrupole phase, there are two different gaps, $\Delta_\perp$ and $\Delta_\parallel$, associated with the out-of-plane and in-plane magnon modes. The out-of-plane gap is finite already within LSWT, and proportional to the deviation from the hidden FM SU(2) point: $\Delta_\perp \propto (1-\lambda)$. The in-plane gap, on the other hand, is zero within LSWT because the classical energy of the vortex pattern is independent of the in-plane rotation angle, $\varphi$. Planar anisotropy $\delta E(\varphi)$ and the corresponding gap only appears beyond LSWT level, via quantum order-from-disorder mechanism, and thus is small: $\Delta_\parallel \propto (1-\lambda)^2$. The octupole phase ($\lambda > 1$) has uniaxial symmetry and a two-fold degenerate magnon dispersion with the gap $\Delta \propto (\lambda - 1)$. As expected, the ordered moments (dashed line) are fully saturated at the hidden FM point. Away from this point, they are reduced by quantum fluctuations (the effect is stronger in the quadrupole phase owing to the presence of a soft in-plane magnon mode).

\subsection{Triangular lattice}
\label{sec:tri}

\begin{figure}[tb]
\begin{center}
\includegraphics[width=0.49\textwidth]{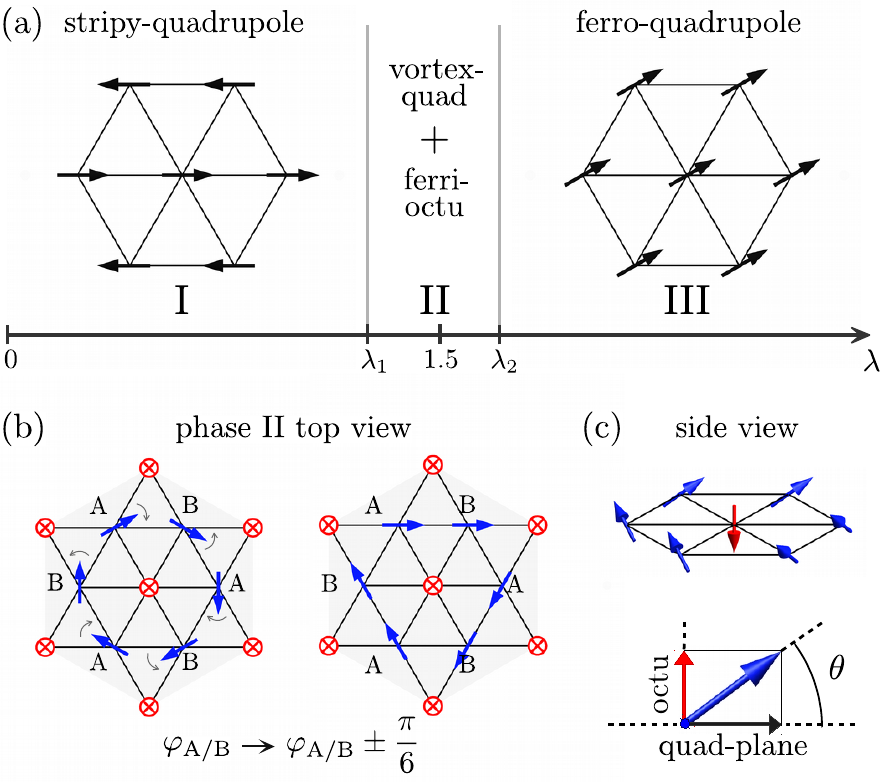}
\caption{
(a)~Phase diagram of Hamiltonian (\ref{eq:KH}) on the triangular lattice as a function of $\lambda=2J/J_\tau$. The phases are separated by spin-flop transitions at $\lambda_1$ and $\lambda_2$. Insets: the ordering patterns of phases I and III. 
(b)~Magnetic unit cell in the quadrupole-octupole mixed phase II. Triangular sublattice (red circles) is occupied by down-oriented octupole moments. On a hexagon, the in-plane components of spins form a vortex. Left and right patterns are related by in-plane $\varphi$ rotations, in opposite directions on $A$ and $B$ sites; quantum effects slightly favor the right pattern.  
(c)~Upper part: down-oriented octupole moment at the center, and out-of-plane canted spins (blue arrows) on a hexagon. Lower part: decomposition of a canted spin into octupole and quadrupole moments.
}\label{fig:tri}
\end{center}
\end{figure}

The behavior of the model (\ref{eq:JtauJ}), or its equivalent (\ref{eq:KH}), on a triangular lattice is of interest too. It is also relevant to double-perovskites, where the face-centered-cubic (fcc) lattice of magnetic ions can be viewed as triangular planes stacked along the [111] direction. The triangular lattice is non-bipartite and a paradigmatic example of geometrical frustration for AF Ising-type models. This is exactly the case for octupolar interactions here: see the $J s^y_is^y_j$ or $\lambda s^y_is^y_j$ terms with positive $J$ and $\lambda$ values in Eqs.~(\ref{eq:JtauJ}) and (\ref{eq:KH}), respectively. 

Leaving full exploration of the model for future study, we discuss now its global phase behavior on a triangular lattice. Inspection of the classical ground states, supported by classical Monte Carlo simulations suggest that there are (at least) three distinct phases, shown in Fig.~\ref{fig:tri}(a), which are realized when the parameter $\lambda=2J/J_\tau$ varies from pure $J_\tau$ limit to a dominant $J$ regime. Stripy-quadrupole phase I at small $\lambda$ is essentially the same state as found earlier in compass model~\cite{Wu08,Che16}; note, however, that the spin pattern in our ``anti-compass'' $\tau$ model is rotated by 90$^\circ$, for the reasons discussed in Sec.~\ref{sec:dd}. This state has a classical energy per site $E_{\rm I}=-(3-\lambda)s^2$ (in units of $J_\tau/2$). 

At large $\lambda$, the ground state is driven by the $J$ interaction, which is of ferro-type in the quadrupolar channel, while octupole AF coupling is equally strong but frustrated. This results in simple ferro-quadrupole order (phase III), with energy $E_{\rm III} =-3(\lambda-1)s^2$. Classically, moments can freely rotate within a quadrupolar plane, but quantum effects generate in-plane anisotropy, pinning the ordered moments along the middle of the two bonds ($\varphi = \frac{\pi}{6}$). The anisotropy is rather weak: magnon zero-point energy, calculated at $\lambda=2$, varies only by $\delta E(\varphi)\simeq 7 \times 10^{-4}J_{\rm \tau}$.

In the above states I and III, the octupolar Ising interaction $\lambda s^y_is^y_j$ was left ``unused'' because of its frustrating nature. At intermediate $\lambda$ values, however, this coupling is actually larger than the quadrupolar one (as $J_\tau$ and $J$ quadrupole terms are of different sign and compete). Therefore, an intermediate state between I and III, which finds a way to resolve ``triangular'' frustration and activates large octupole couplings, is expected.

The pseudospin ordering pattern, whose unit cell is shown in Fig.~\ref{fig:tri}(b) and further detailed in Fig.~\ref{fig:tri}(c), does exactly this job. In this state, the original triangular lattice is divided into two, honeycomb and triangular sublattices. The spins on the honeycomb sublattice are canted and carry both octupole and quadrupole moments. The latter condense into a vortex pattern similar to what shown in Fig.~\ref{fig:honey}(a). While the out-of-plane components form a ferro-octupole order. The honeycomb octupole moment is largely (but not fully) compensated by down-oriented octupoles residing at the middle of every hexagon. As a whole, phase II represents a coherent superposition of vortex-quadrupole and ferri-octupole orders. Such a mixture of the different rank multipoles is rather unusual. We also note that this order is noncoplanar and has a large unit cell which helps to relieve the frustrations inherent to spin-orbital models. In this sense, the case is similar to a complex behavior of spin-orbit pseudospins $J=1/2$ of $d^5$ ions on a triangular lattice~\cite{Kha05,Rou16,Cat15}. 

As a function of spin canting angle $\theta$, the classicial energy of the mixed state II is obtained as follows:  
 \begin{equation}
E_{\rm II}(\theta)/s^2 = -1 - 2\lambda\sin{\theta} + (1+\lambda)\sin^2{\theta}.
 \end{equation}
Here, the second term originates from coupling between honeycomb lattice octupoles $s^y\propto\sin{\theta}$ with those residing at the hexagon centers. Minimizing $E_{\rm II}(\theta)$ with respect to $\theta$, we find $\sin{\theta} = \lambda/(1+\lambda)$. This gives a ground state energy of phase II (per site):    
\begin{equation}
 E_{\rm II}= -\left(\lambda + \frac{1}{1+\lambda}\right)s^2. 
 \end{equation}
Comparing this result with $E_{\rm I}$ and $E_{\rm III}$ obtained above, we find the phase transition points $\lambda_1$ and $\lambda_2$:
\begin{equation}
 \lambda_1 =\frac{1+\sqrt{17}}{4} \simeq 1.28, \;\;\;\;\;
 \lambda_2 =\frac{1+\sqrt{33}}{4} \simeq 1.69. 
 \end{equation}

In between $\lambda_1$ and $\lambda_2$, the angle $\theta$ varies from 34$^\circ$ to 39$^\circ$, and the size of octupole moment on honeycomb sites $2|\langle s^y \rangle|=\sin{\theta} \simeq 0.56-0.63$. This nearly compensates pure octupole moments from triangular sublattice, leaving rather small total octupole moment per site: $2|\langle s^y \rangle|_{tot}=\tfrac{1}{3}(2\sin{\theta}-1)\simeq 0.04 - 0.09$.

Displayed in Fig.~\ref{fig:tri}(b) are two different vortex patterns, related to each other by in-plane rotations of spins. Classically, these states are degenerate. Quantum zero-point energies, calculated within LSWT for these two ground states, slightly differ; the right one is lower by $\delta E \simeq 2 \times 10^{-4} J_{\rm\tau}$. This result implies that in-plane magnon excitations acquire a small but finite gap, generated by the order-from-disorder mechanism. Out-of-plane excitations, corresponding to fluctuations between quadrupole and octupole sectors, are gapped out already on a classical level. 

Overall, the $E_g$ pseudospin $J_\tau-J$ model (\ref{eq:JtauJ}) on triangular lattice contains rich physics yet to be fully explored theoretically. The above results also should encourage experimental work finding and studying 5$d^2$ compounds with quasi-two dimensional honeycomb and triangular lattice structures.    

\subsection{Double perovskites}
\label{sec:DP}

Now, we move to the DP lattice which motivated this study. DPs are special because the magnetic ions are widely separated from each other and thus interact weakly. This implies that the pseudospin one-half description, which assumes that the intersite interactions are less than on-site cubic splitting $\Delta_c$, is best justified in DP compounds. 

The dominant exchange channel in DPs is due to the single-orbital process considered in Sec.~\ref{sec:dd}. This results in Kugel-Khomskii type Hamiltonian (\ref{eq:tau}) acting in the pure quadrupole $\tau$-channel: $J_\tau  \tau_{i\gamma}\tau_{j\gamma}$. To our knowledge, the ground state of this model on fcc lattice (formed by magnetic ions in DPs) has not yet been considered. To address the behavior of the $\tau$-model on the highly frustrated fcc lattice, we perform classical Monte-Carlo simulations on the related model, $J_\tau  n_{i\gamma}n_{j\gamma}$, where pseudospins $\tau=1/2$ are replaced by classical vectors of unit length ($\vc n^2=1$). 

\begin{figure}[tb]
\begin{center}
\includegraphics[width=0.45\textwidth]{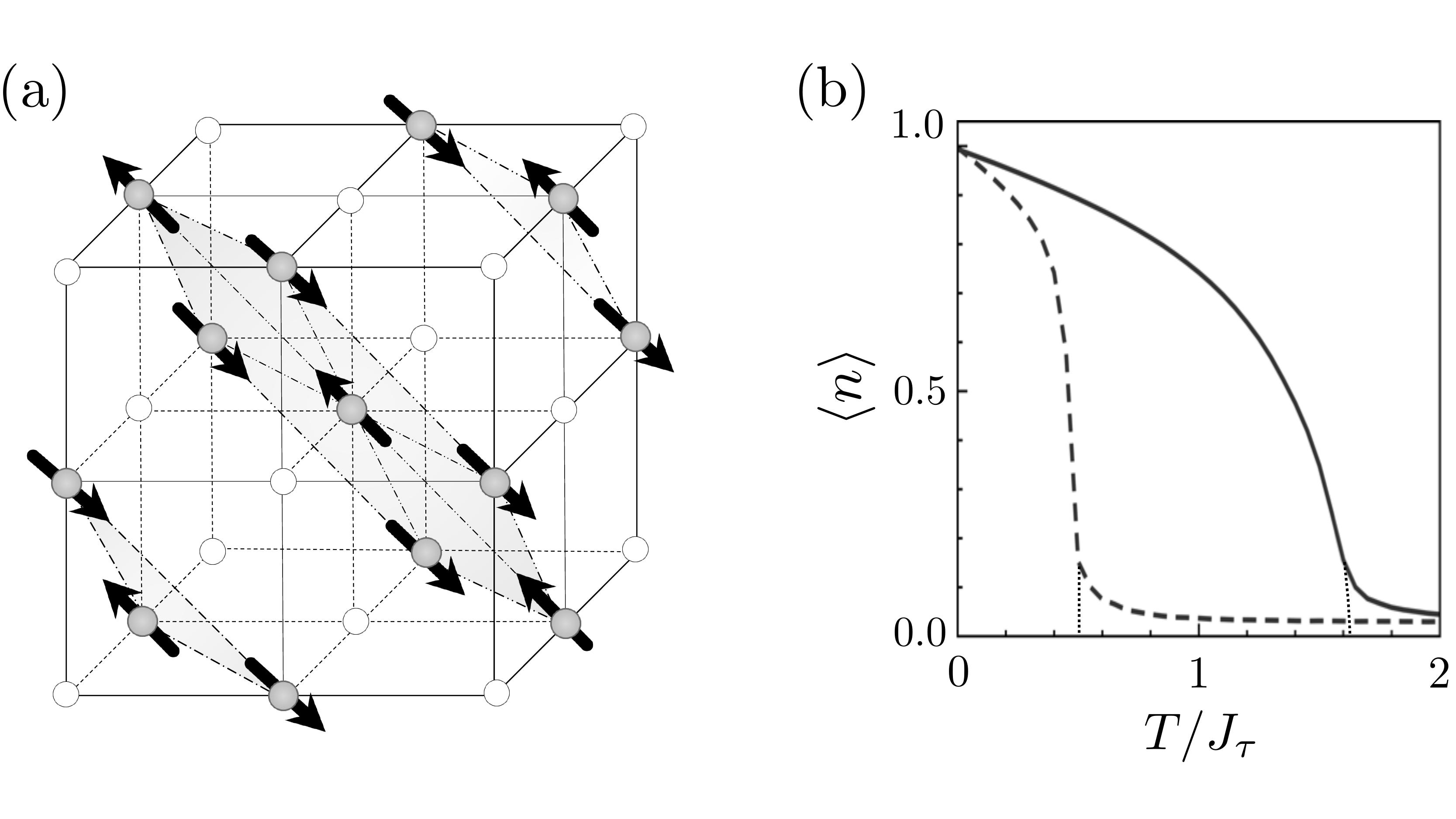}
\caption{
(a)~Double perovskite lattice, where the transition metal ions (filled circles) reside on the fcc sublattice. Arrows show the ordered pattern of quadrupole moments. It can be viewed as a stack of triangular lattice [111] planes (shaded), with stripy-AF quadrupole order within each plane. 
(b)~Temperature dependence of the order parameter $\langle n \rangle$ for classical $J_\tau$ model in DP (solid line) and triangular (dashed line) lattices, obtained by classical Monte Carlo simulation.
}\label{fig:DP}
\end{center}
\end{figure}

The simulated annealing Monte Carlo is performed for DP and, for comparison, triangular lattices. We use 1372 sites (7 $\times$ 7 $\times$ 7 unit cell) of DP and 1296 sites ($36 \times 36$) of triangular lattice with periodic boundary conditions. Monte Carlo simulations were performed using the ALPS project library \cite{alps1,alps2,alps3}. We find a collinear AF-quadrupole order at low temperatures, and the ordering pattern in DP lattice is displayed in Fig.~\ref{fig:DP}(a). The moment is along the bond direction, and there are 8 anti-parallel and 4 parallel nearest-neighbors. Within the [111] planes, moments form a stripy pattern as in the triangular lattice, see phase I in Fig.~\ref{fig:tri}(a). Temperature dependence of the ordered moment length $\langle n \rangle$ in Fig.~\ref{fig:DP}(b) shows that the ordering sets in at $T_c \simeq 1.6 J_\tau$ in DP lattice. This is about three times higher than $T_c/J_\tau$ in the triangular lattice, most likely due to increased dimensionality. Quantifying the quadrupole moment reduction by quantum fluctuations in fcc lattice is an interesting but challenging problem, and left for future study. 

It should also be noticed that $T_c$ in the actual model with quantum spin $\tau=1/2$ is different from the above Monte Carlo result. Roughly, an upper limit for the rescaling factor can be obtained by replacing $\vc n^2=1$ by $s(s+1)$. For the present case of spin one-half, this gives an estimate of $T_c \sim J_\tau$. A rather low value of $T_c$ (despite a large coordination number 12) is presumably due to frustrations of the model on fcc lattice. 

In principle, the 90$^\circ$ bonding superexchange via nonmagnetic $B'$ sites is possible in DP lattice, due to the extended nature of 5$d$ orbitals. However, this process involves many hopping steps ($B_i$-$O$-$B'$-$O$-$B_j$), and the corresponding indirect hopping $t$ and hence $J$ must be small. Therefore, even though the full exchange Hamiltonian in DPs is formally given by Eq.~(\ref{eq:JtauJ}), the single-orbital quadrupole interaction $J_\tau$ should dominate over $J$ coupling between the octupoles. This implies that spin-orbital exchange in DPs uniquely supports AF order of quadrupole moments, breaking underlying discrete point group symmetries both in real and pseudospin spaces, but preserves TR symmetry.   

We should note that the above result is obtained at large cubic splitting $\Delta_c$ between the excited $T_{2g}$ states and pseudospin $E_g$ doublet. The previous calculations in the other limit, i.e. neglecting cubic splitting and using full $J=2$ Hilbert space instead \cite{Che11} found no octupolar instability, too. Ref.~\cite{Par20} suggested that in the intermediate case, when the $E_g$ doublet is formed and $T_{2g}$ triplet is not too high, the virtual states may generate octupolar interactions that are strong enough to overcome quadrupolar couplings. We now inspect this possibility, by considering contributions of the virtual $T_{2g}$ states to the effective pseudospin $E_g$ Hamiltonian. This brings us to Jahn-Teller physics which operates not only within the ground state $E_g$ doublet, but also connects it with the $T_{2g}$ triplet. 

\section{Jahn-Teller coupling effects. Implications for double-perovskites}
\label{sec:JT}

\subsection{Jahn-Teller Hamiltonian}
\label{sec:JTH}

We consider a linear coupling between the octahedral normal modes $Q_\Gamma$ and electron quadrupolar moments $O_\Gamma$ of symmetry $\Gamma$. For $t_{2g}$ orbital systems, the $\Gamma_3$ doublet ($Q_3$ and $Q_2$ modes of $3z^2-r^2$ and $x^2-y^2$ symmetries, respectively) as well as $\Gamma_5$ triplet modes ($Q_{xy}$, etc) are relevant \cite{Abr70}. Microscopically, orbital-lattice coupling in the $\Gamma_3$ channel splits the $t_{2g}$ orbital levels, while coupling to $\Gamma_5$ modes is orbital non-diagonal, e.g. $Q_{xy}$ distortion mixes $d_{yz}$ and $d_{zx}$ wavefunctions.

In terms of two-electron $J=2$ quadrupoles, the Jahn-Teller couplings in the above two channels read as follows (summation over lattice sites $i$ is implied):    
\begin{align}
\label{eq:G3}
\mathcal{H}_{JT}(\Gamma_3) &= - g\;(Q_3O_3 +Q_2O_2)_i , \\ 
\mathcal{H}_{JT}(\Gamma_5) &= - g'(Q_{xy}O_{xy}+Q_{yz}O_{yz}+Q_{zx}O_{zx})_i ,   
\label{eq:G5}
\end{align}
where $O_3$ and $O_2$ quadrupoles are defined in Eqs.~(\ref{eq:O3}) and (\ref{eq:O2}), while $O_{xy}=(J_x J_y + J_y J_x)/2\sqrt 3$, etc. Within the ground state $E_g$ doublet, $\Gamma_3$ coupling takes a form familiar from the $e_g$ orbital JT problem \cite{Kan60,Kug82}:  
\begin{equation}
  \mathcal{H}_{JT}(\Gamma_3)=-g\;(Q_3\sigma^z +Q_2\sigma^x),
  \;\;\;\;\; \sigma^{z/x}=2s^{z/x}. 
\label{eq:JTEg}
\end{equation}
On the other hand, quadrupolar operators in $\mathcal{H}_{JT}(\Gamma_5)$ have no matrix elements within the pseudospin subspace; instead, they connect the $E_g$ doublet to the excited $T_{2g}$ states. This leads to a so-called second-order or ``pseudo-Jahn-Teller'' effect \cite{Ber06} which operates through the mixing of the ground and excited states. In spin-orbit coupled systems, this effect modulates the spatial shape of the pseudospin wavefunctions and generates new terms in low-energy effective Hamiltonians \cite{Liu19}.

\subsection{Pseudospin interactions mediated by Jahn-Teller coupling}
\label{sec:int}

Spatial correlations between the octahedral deformations on different sites mediate interactions between quadrupolar moments \cite{Kan60,Geh75}. Typically, these interactions cooperate with the Kugel-Khomskii mechanism of orbital ordering \cite{Kug82}. In most TM compounds, the JT centers share common oxygens and thus stay in direct contact with each other. This leaves little room for single-ion JT dynamics. In the DP lattice, however, the JT ions have no common oxygen, so they have to interact by exchanging virtual phonons. Since JT phonon modes disperse weakly, this interaction is much smaller than in perovskites. This has two important consequences: (i) cooperative JT couplings are weak enough so that the $E_g$ pseudospin description remains valid, and (ii) single-ion JT dynamics, intrinsic to non-Kramers $E_g$ states, may develop.

\begin{figure}[tb]
\begin{center}
\includegraphics[width=0.4\textwidth]{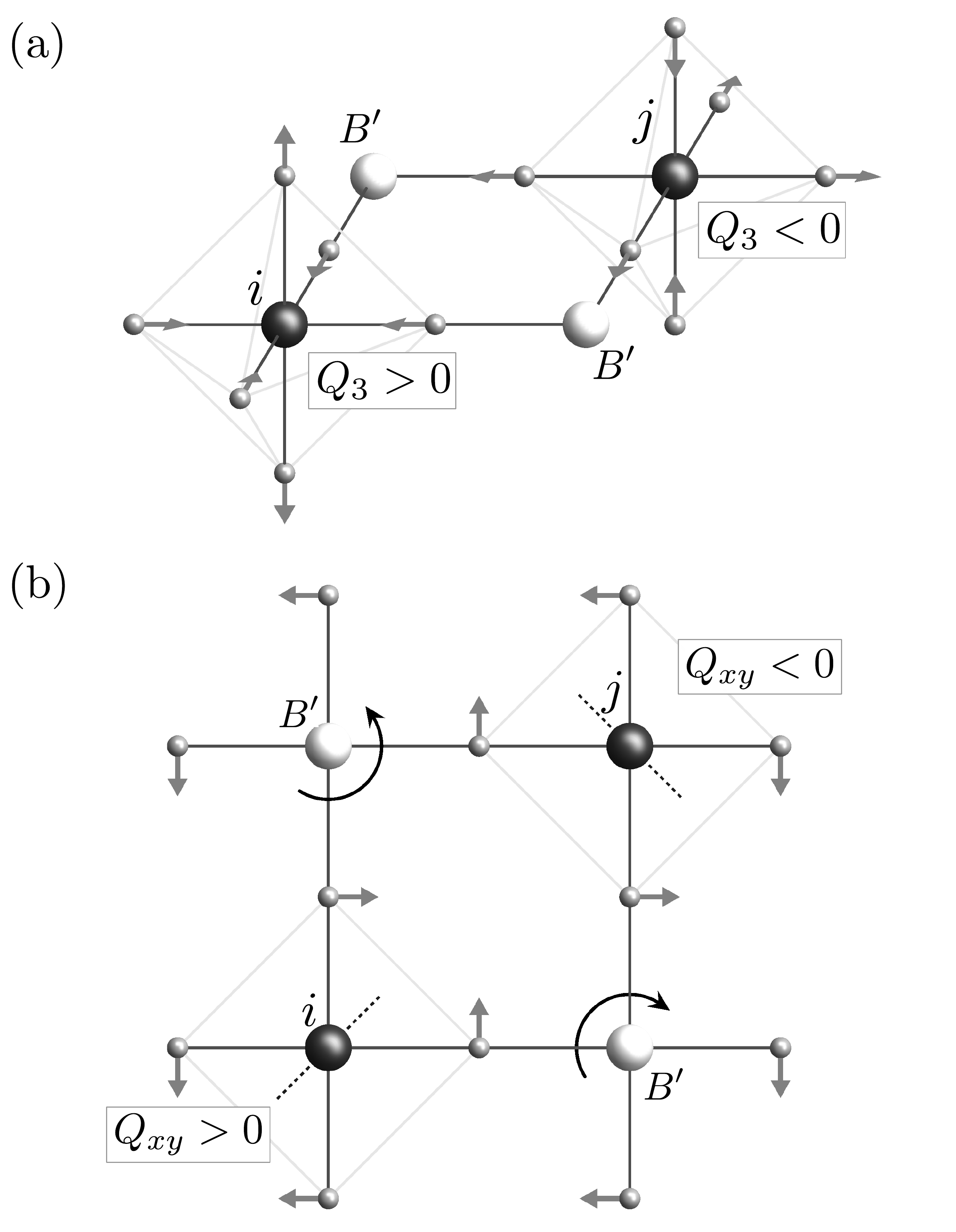}
\caption{
(a)~Anti-ferro correlated $Q_3$ distortions on $xy$ plane of the DP lattice, leading to $V$ coupling (\ref{eq:V}) between quadrupole moments of $\Gamma_3$ symmetry. 
(b)~Anti-ferro correlated $Q_{xy}$ distortions on $xy$ plane, leading to $V'$ coupling (\ref{eq:V'}) in the $\Gamma_5$ quadrupolar channel.   
}\label{fig:JT}
\end{center}
\end{figure}

We start with pseudospin interactions mediated by JT coupling in the $\Gamma_3$ channel (\ref{eq:JTEg}). Two ions in the $xy$ plane couple most efficiently via $Q_3$ type distortions, as illustrated in Fig.~5(a). The corresponding quadrupole interaction is 
\begin{equation}
V\sigma^z_i\sigma^z_j=4Vs^z_is^z_j \;,
\label{eq:V}
\end{equation}
whose strength $V=-g^2\langle Q_{3i} Q_{3j}\rangle_{\omega=0} >0$ is given by non-local static susceptibility of the $Q_3$ modes. In DPs, antiferro-type intersite correlations $\langle Q_{3i} Q_{3j}\rangle<0$ arise due to a finite dispersion $\delta \omega_q$ of the corresponding optical phonons with energy $\omega_0$. This gives a rough scale of $V$ as a small fraction $\propto \delta \omega_q/\omega_0 \sim 0.1$ of a single-ion JT stabilization energy $E_{JT}$. For $d^1$ Os in DPs, Refs.~\cite{Iwa18,Xu16} evaluated $E_{JT}\sim 20$~meV (this might be lower for $d^2$), suggesting $V\sim 1$~meV, i.e. of the same order as the exchange coupling $J_\tau$ for DPs. Relatively small values of $E_{JT}$ (and hence JT mediated coupling $V$) may indicate a rather weak coupling of the diffuse 5$d$ orbitals to the lattice. We note that considering both $Q_3$ and $Q_2$ modes would induce also $s^x_is^x_j$ and $s^z_is^x_j$ type terms; however, these are not essential in the context of possible octupole order. The main result is that the quadrupole interactions, mediated by $\Gamma_3$ type JT phonons cooperate with the exchange coupling, i.e. $J_\tau \rightarrow J_\tau + 4V$, as in the case of the usual $e_g$ orbital problem in manganites.  

The JT coupling in the $\Gamma_5$ channel (\ref{eq:G5}) works differently, and it actually leads to octupolar coupling $s^y_is^y_j$. For a pair in the $xy$ plane, we consider that $Q_{xy}$ type distortions shown in Fig.~5(b) are most relevant. This leads to the following interaction between $\Gamma_5$-type quadrupole moments: 
\begin{equation}
V' \; O_{xy}^i O_{xy}^j \;,
\label{eq:V'}
\end{equation}
with positive $V'=-(g')^2\langle Q_{xy}^i Q_{xy}^j\rangle_{\omega=0}$. The interactions on $yz$ and $zx$ planes are induced by $Q_{yz}$ and $Q_{zx}$ type modes. For $t_{2g}$ orbital systems, constants $V$ and $V'$ are expected to be of the same order, but their ratio depends on material details. In $d^1$ Os double-perovskite, coupling to $\Gamma_3$ type modes are stronger~\cite{Iwa18} which might also be the case in $d^2$ osmates.   

As said above, $\Gamma_5$ quadrupoles have no matrix elements within $E_g$ doublet; instead, they create transitions from the $E_g$ doublet to excited $T_{2g}$ states. For example, $O_{xy}=i(T^\dagger_zf_\downarrow-f^\dagger_\downarrow T_z)$, where hard-core bosons $f$ and $T$ belong to pseudospin $(f_\uparrow,f_\downarrow)$ and triplet $(T_x,T_y,T_z)$ sectors, correspondingly. Thus, the pairwise interaction $V'O_{xy}^i O_{xy}^j$ may (a) excite a pair of triplons, and (b) lead to dispersion and broadening of the $T$-excitons. In the present context, we are interested in the pair generation process which dynamically mixes up the ground and excited state wavefunctions, modifying thereby the pseudospin exchange Hamiltonian. Specifically, this process activates the composite operators $(\vc{S}L_z^2)_i$ and $(\vc{S}L_z^2)_j$ in the exchange Hamiltonian of Eq.~(\ref{eq:Hdd}). These operators have nondiagonal $E_g\leftrightarrow T_{2g}$ matrix elements [e.g., $S_zL_z^2=-(T^\dagger_zf_\uparrow+H.c.)$], and are thus sensitive to the admixture of JT induced triplet states in pseudospin wavefunctions. 

Now, we assume that the exchange $J_\tau$ and quadrupole $V'$ couplings are small compared to cubic splitting $\Delta_c$. In other words, we assume that the dispersion and broadening of triplon excitations, caused by these interactions, is smaller than $\Delta_c$, and thus the pseudospin description is valid. This is exactly what is observed in experiment \cite{Mah20}. Then we proceed along the lines of Ref.~\cite{Par20}, eliminating virtual triplon pairs perturbatively. This results in the following effective Hamiltonian (for pairs on the $xy$ plane), which now includes the exchange as well as JT-coupling mediated interactions:
\begin{equation}
  \mathcal{H}_{eff}^{(z)}=J_z s_i^zs_j^z + J_x s_i^xs_j^x - J_y s_i^ys_j^y. 
\label{eq:final}
\end{equation}
For $\gamma=x,y$ bonds, one has to replace $s^z \rightarrow \tau_\gamma$ and $s^x \rightarrow \bar\tau_\gamma$. Effective parameters read as:
\begin{align}
\label{eq:Jz}
J_z &= J_\tau (1-\frac{81}{32}\frac{J_\tau}{\Delta_c}) +
4V (1-\frac{V'}{8V}\frac{V'}{\Delta_c}) , \\ 
J_y &= J_x=  J_\tau \;\frac{9V'}{4\Delta_c}\;.  
\label{eq:Jy}
\end{align}
Provided that the triplet excitations are well separated from the pseudospin doublet, i.e. $J_\tau$ and $V'$ much smaller than $\Delta_c$, the corrections $\propto 1/\Delta_c$ to $J_z$ in Eq.~(\ref{eq:Jz}) can be neglected, and we obtain 
\begin{equation}
\frac{J_y}{J_z} \simeq\frac{J_\tau}{J_\tau+4V}\; \frac{2V'}{\Delta_c} < 1.
\label{eq:ratio}
\end{equation}
This implies that quadrupolar $\tau$-interactions dominate over those in the octupolar $s^y$ sector. Moreover, $J_y/J_z$ can be reduced by dynamical JT effects, which suppresses the interactions involving $s^y$ operators by the Ham factor~\cite{Abr70}. The result (\ref{eq:ratio}) is natural for 5$d$ ions on DP lattices, where intersite interactions $J_\tau$, $V$, and $V'$ between widely separated ions should be much less than the single-ion energy $\Delta_c$. The latter is driven by large SOC for 5$d$ electrons; we also think that the cubic splitting $\Delta_c$ is further enhanced by the dynamical JT effect stabilizing the $E_g$ doublet against the $T_{2g}$ states. It would be interesting to check the latter point by quantum chemistry calculations.   

The above findings suggest that a single phase transition observed in 5$d^2$ osmium DP oxides \cite{Mah20} is driven by a quadrupolar ordering of $E_g$ doublets. Comparison of the Monte-Carlo result (Sec.~\ref{sec:DP}) for the quadrupolar ordering temperature $\sim J_\tau +4V$ (including now JT-phonon mediated coupling $V$) with the experimental transition temperature $30-50$~K gives $J_\tau +4V \sim 4$~meV, consistent with the above estimates of intersite couplings. Concomitant lattice distortions might be small for several reasons: first, JT coupling is already weakened by the SOC effect which partially removes orbital degeneracy; second, on-site JT vibronic dynamics~\cite{Abr70,Ber06} and intersite quantum fluctuations reduce the pseudospin order parameter and hence the static distortions, to the levels that are difficult to detect directly by x-ray diffraction. In general, JT coupling seems to be rather weak for 5$d$ orbitals; indeed, quadrupole order induced distortions in 5$d^1$ DPs have been found to be extremely small~\cite{Hir20} or below the resolution limit~\cite{Hir21}. However, quadrupolar order should lead to changes in phonon spectra that should be well visible in Raman and optical data. Possible signatures of the dynamical JT effect, e.g. transitions between vibronic levels \cite{Guh75} are of special interest. Also, a quadrupolar order can be probed by nuclear magnetic and quadrupole resonance experiments. Further experiments, especially on single crystal samples, are necessary to identify the nature of the "hidden" order parameter in DP 5$d^2$ compounds. 

\subsection{Coupling to local distortions: Induced magnetic moments}
\label{sec:imp}

While a quadrupole order of JT active $E_g$ doublets sounds natural, this picture cannot explain TR symmetry breaking observed in 5$d^2$ DPs \cite{Tho14,Mar16,Aha10}. One possible explanation is to attribute this effect to magnetic moments induced by defects (e.g., $B\leftrightarrow B'$ site disorder). Near defects, non-cubic crystal fields can modify the $E_g$ doublet wavefunctions, or even completely destroy the pseudospin description, thus recovering the $\vc J$ dipole moments at least partially. Nuclear magnetic and quadrupole resonance lineshapes may quantify such magnetic state inhomogeneities. In fact, the signatures of spin disorder and freezing are rather common in $d^2$ DPs \cite{Aha10,Tho14,Mar16}. To show how symmetry lowering distortions affect the physical content of the $E_g$ doublets and their interactions, we consider JT coupling of $\Gamma_5$ symmetry quadrupoles $O_{\alpha\beta}$ to the corresponding local distortions $e_{\alpha\beta}$:
\begin{equation}
\mathcal{\delta H}_i= - g'(e_{xy}O_{xy}+e_{yz}O_{yz}+e_{zx}O_{zx})_i \;. 
\label{eq:loc}
\end{equation}
The local quadrupolar fields (\ref{eq:loc}) modify the $E_g$ doublet functions at site $i$ as follows: 
\begin{align}
\label{eq:tup}
|\tilde\uparrow\rangle_i & \Rightarrow |\!\uparrow\rangle+ is(\delta_x|T_x\rangle-\delta_y|T_y\rangle)_i, \\ 
|\tilde\downarrow\rangle_i & \Rightarrow |\!\downarrow\rangle+
i(\delta_z|T_z\rangle-c\delta_x|T_x\rangle-c\delta_y|T_y\rangle)_i, 
\label{eq:tdown}
\end{align}
where $s=\sqrt 3/2$, $c=1/2$ [the normalization factors $p_\uparrow^2=1+s^2(\delta_x^2+\delta_y^2)$ and $p_\downarrow^2=1+\delta_z^2 + c^2(\delta_x^2+\delta_y^2)$ are not shown]. The parameters $\delta_z=g'e_{xy}/\Delta_c$, etc quantify the degree of admixture of virtual triplet states $T_x, T_y$, and $T_z$ into the ground state due to strain $e_{\alpha\beta}$ field. This admixture ``magnetizes'' the $E_g$ doublet, by inducing a dipolar component into the $s^y$ operator. By calculating matrix elements of total angular momentum $\vc J$ within the modified $E_g$ doublet (\ref{eq:tup}, \ref{eq:tdown}), we find an induced moment $J_{i\alpha}=4\delta_{i\alpha} s^y_i$, illustrating a partial recovery of the dipolar moments due to local distortions. The corresponding magnetic moment, which is carried by the $s^y$ operator, is $M_{i\alpha}=2\delta_{i\alpha} s^y_i$ (using $g$ factor $g=1/2$ of $J=2$ state).

In principle, a direct link between lattice distortions and magnetism is generic to all spin-orbit Mott insulators. In 5$d^2$ ion systems, where the non-magnetic nature of the $E_g$ doublet is protected by cubic symmetry (i.e. independent of covalency, etc), lattice distortions have an especially strong impact on magnetic properties. To illustrate this point further, we may consider uniform strain applied along the [111] axis of a crystal: $e_{xy}=e_{yz}=e_{zx}=e/\sqrt 3$. This induces a magnetic moment with $g$ factor $g_\parallel=\frac{2}{\sqrt 3}\frac{\Delta_{tr}}{\Delta_c}$, where $\Delta_{tr}=g'e$ is the strain induced field, while the $g$ factors within the [111] plane remain zero. This results in an extreme anisotropy of the magnetic response to lattice distortions in 5$d^2$ systems. 
  
Non-cubic crystal fields also induce new couplings between pseudospin moments. Projection of the exchange interaction (\ref{eq:Hdd}) onto the $E_g$ doublet with ``distorted'' wavefunctions (\ref{eq:tup}, \ref{eq:tdown}) modifies the $J_\tau$ term in Eqs. (\ref{eq:tau}) and (\ref{eq:JtauJ}) as follows:
\begin{equation}
  J_\tau\; \tau_{i\gamma}\tau_{j\gamma} \rightarrow
  J_\tau \left[\tau_{i\gamma}\tau_{j\gamma} + a_{ij}^{(\gamma)} s^y_is^y_j\right],  
\label{eq:JdJT}
\end{equation}
where a new term, the $s^y_is^y_j$ coupling which operates in the magnetic channel, appears. Its relative strength is given by $a_{ij}^{(z)}=9(\delta_{iz} \delta_{jz} +\tfrac{1}{4}\delta_{ix} \delta_{jx} +\tfrac{1}{4}\delta_{iy} \delta_{jy})$ for $z$ type bonds (results for $\gamma=x, y$ follow from symmetry). We see that even rather small strain fields are sufficient to support $s^y$ order locally: the new term becomes comparable with quadrupolar $\tau$-coupling already at $\delta \sim 1/3$. The sign of $a_{ij}$ depends on the relative orientation of local distortions; for antiferro-type distortions ($\delta_i \delta_j<0$), the $s^y$ moments are coupled ferromagnetically, and vice versa (i.e. following Goodenough-Kanamori rules).    

The lattice effects on the physical content of the $E_g$ doublets should be essential for understanding the magnetic properties of 5$d^2$ osmates. A qualitative picture is that while pseudospin one-half ordering in these compounds is predominantly of a quadrupolar type, there should also be a weak and spatially inhomogeneous dipolar component of the pseudospin order parameter, induced by the random distortions inevitable in real crystals. 


\section{Conclusions}
\label{sec:Con}

We have developed a microscopic theory for multipole orders in spin-orbit Mott insulators of non-Kramers $d^2$ ions, which have a non-magnetic doublet ground state of $E_g$ symmetry. The exchange Hamiltonians for various hopping processes are derived and expressed in terms of pseudospin-1/2 operators. Reflecting the spin-orbital mixed nature of the $E_g$ wavefunctions, the pseudospin interactions are in general strongly anisotropic and depend on the bond directions. The phase behavior of these models on different lattices are considered by means of analytical and numerical methods.

On a honeycomb lattice, we find that interplay between direct overlap of $t_{2g}$ orbitals and their hopping via ligand ions gives rise to a competition between two distinct multipole orders: vortex-type quadrupole order and collinear AF octupole order. These two states become degenerate at the parameter point where the model has a hidden SU(2) symmetry and is isomorphic to the FM Heisenberg model. The model also can be mapped to the extended Kitaev model for $d^5$ systems, which is useful to understand its global phase behavior.

On a triangular lattice, a combination of geometrical and spin-orbital frustrations result in a novel type of ordering which can be viewed as a coherent superposition of vortex-type quadrupole and ferri-type octupole orders. This complex state appears as an intermediate phase between collinear AF and FM quadrupole states, as a compromise to their competition. 

Double perovskite compounds of $d^2$ ions with strong SOC such as osmium Os$^{6+}$ or rhenium Re$^{5+}$ are discussed in more detail, including also Jahn-Teller coupling of electron quadrupole moments to lattice degrees of freedom. We find that the exchange and JT effects do work cooperatively to support quadrupole order in DP lattices. Static lattice distortions associated with this order are expected to be small, due to a reduction of the order parameters by the dynamical JT effect and pseudospin frustrations caused by bond-dependent interactions on the fcc lattice. Nevertheless, quadrupole order should lead to well detectable changes in phonon spectra as well as in nuclear magnetic resonance lineshapes. Signatures of JT dynamics in Raman and optical data should also be interesting to look for. A possible scenario for TR symmetry breaking due to noncubic crystal fields near defects is suggested. We show that the local distortions modify the ground state wavefunctions, induce local magnetic moments and enhance their exchange interactions, illustrating the importance of the lattice effects for interpretation of the experimental data.

Overall, the models introduced and discussed in this paper are of interest in their own right. From a materials perspective, our findings suggest rich physics yet to be explored in 5$d^2$ spin-orbit Mott insulators. The present work forms a theoretical basis for the future research of these compounds.


\acknowledgments

We would like to thank J. Chaloupka, B. D. Gaulin, and A. Paramekanti for useful discussions. G.Kh. acknowledges support by the European Research Council under Advanced Grant No. 669550 (Com4Com). H.Y.K. acknowledges support from the NSERC Discovery Grant No. 06089-2016, and support from CIFAR and the Canada Research Chairs Program. Computations were performed on the Niagara supercomputer at the SciNet HPC Consortium. SciNet is funded by: the Canada Foundation for Innovation under the auspices of Compute Canada; the Government of Ontario; Ontario Research Fund - Research Excellence; and the University of Toronto.



\begin{thebibliography}{99}
\bibitem{note_G3} We note that the notations $E_g/e_g$ and $T_{2g}/t_{2g}$ correspond to $\Gamma_3$ and $\Gamma_5$ representations of the cubic group, respectively.
\end{thebibliography}
\end{document}